\documentclass[aps,prx,twocolumn,superscriptaddress,longbibliography]{revtex4-2}
\usepackage[utf8]{inputenc}
\usepackage{amsmath}
\usepackage{amssymb}
\usepackage{epsfig}
\usepackage{epstopdf}
\usepackage{graphicx,url}
\usepackage{bm}
\usepackage{mathrsfs}
\usepackage{tikz}
\usepackage{comment}
\usepackage{physics}
\usepackage{appendix}
\usepackage{hyperref}
\usepackage{comment}
\usepackage{natbib}

\newcommand{\adam}[1]{\textcolor{red}{#1}}

\begin{document}

\title{Collective chemotactic search}

\author{Adam Wysocki}
\email{a.wysocki@lusi.uni-sb.de}
\affiliation{Department of Theoretical Physics \& Center for Biophysics, Saarland University, 66123 Saarbrücken, Germany}

\author{Hugues Meyer}
\email{hugues.meyer@nottingham.ac.uk}
\affiliation{School of Physics and Astronomy, University of Nottingham, Nottingham NG7 2RD, United Kingdom}
\affiliation{Department of Theoretical Physics \& Center for Biophysics, Saarland University, 66123 Saarbrücken, Germany}

\author{Heiko Rieger}
\email{heiko.rieger@uni-saarland.de}
\affiliation{Department of Theoretical Physics \& Center for Biophysics, Saarland University, 66123 Saarbrücken, Germany}

\date{\today}

\begin{abstract}
We investigate collective search by self-propelled agents that are repelled by their own chemically produced trails, a minimal mechanism that simultaneously generates indirect interactions and memory. Using lattice and off-lattice models, we show that this mechanism enhances search efficiency through two distinct regimes. In a weak-memory regime, chemical cues are short-lived and interactions primarily promote spatial separation between agents. This reduces redundant exploration while preserving mobility, leading to an optimal trade-off between spatial order and persistence. In a strong-memory regime, long-lived chemical trails induce effective self-avoidance, strongly suppressing revisits and long search times. Here optimal search occurs at finite memory strength: permanently persistent trails lead to self-caging, while moderate memory enables efficient exploration. At higher densities, overlapping chemical trails give rise to a collective self-avoidance mechanism that yields substantial cooperative speedup without global spatial order. Together, these results show how chemically mediated memory and interactions can optimize collective search across distinct dynamical regimes.
\end{abstract}

\maketitle

\section{Introduction}

Search processes, in which mobile agents explore an environment to locate targets, are ubiquitous across physics, biology, and robotics
\cite{targetproblem}. 
Physically relevant search processes 
are often stochastic in nature \cite{vankampen,redner,bressloff}
and include chemical reaction kinetics \cite{smoluchowski,collins,rice}, genetic transcription \cite{ogberg,gorman2008visualizing}, bacteria searching for nutrients \cite{berg} or immune cells searching for pathogenic cells \cite{krummel2016t}, foraging animals 
\cite{hassell1978foraging, traniello1989foraging, edwards2007revisiting,hills2013adaptive}
or swarming robots \cite{senanayake2016search, drew2021multi}, to name but a few. In all these settings, search efficiency is a key performance metric and is commonly quantified by the mean first-passage time (MFPT) \cite{redner}, the average time required to reach a target for the first time.

The problem of optimizing search has been studied extensively for single agents and for collections of non-interacting searchers. Optimal strategies are known for a wide range of stochastic processes, including 
Lévy flights \cite{lomholt2008levy}, 
persistent \cite{tejedor2012optimizing}
and intermittent random walks \cite{benichou2011intermittent},
random walks with $n$-step memory \cite{meyer2021optimal,klimek2022optimal}, 
random walks with resetting \cite{kusmierz2014first, chechkin2018random, bressloff2020search, stanislavsky2021optimal, tal2020experimental}, and random walks interacting with the environment \cite{altshuler2023environmental}.

For multiple {\it independent} (i.e. non-interacting) 
agents, the MFPT typically decreases trivially with agent number, and the optimal strategy largely mirrors that of a single searcher \cite{rednerkaprivsky,oshanin,mejia2011first,nayak,grebenkov1,grebenkov2}. By contrast, far less is known about how \emph{interactions} between agents can qualitatively alter collective search efficiency
\cite{tani2014optimal,ro2023pre,martinez2013optimizing}.

A central limitation of {\it collective search} is redundant exploration. Repeated visits to already explored regions slow down target discovery and diminish the benefits of parallelism. In principle, redundant exploration can be suppressed in two complementary ways. First, agents may retain \emph{memory} of their past trajectories and avoid revisits. Second, interactions between agents may promote spatial separation, allowing the search domain to be effectively divided into non-overlapping regions \cite{tani2014optimal,ro2023pre,martinez2013optimizing}. While memory is known to strongly enhance search efficiency at the single-agent level \cite{meyer2021optimal,SpatialMemory,klimek2022optimal}, its role in collective search—and its interplay with inter-agent interactions—remains poorly understood.

In this work, we address these questions using auto-chemotactic particles \cite{taktikos2011pre,liebchen1,kranz2019trail,mokhtari2022spontaneous,barbier2022self,meyer2023alignment} as a minimal and universal model of collective search. 
Auto-chemotactic agents produce and respond to a chemical field, thereby interacting both with their own past trajectories and with chemical traces left by other agents. This single mechanism naturally combines \emph{self-interaction}, which implements memory, with \emph{indirect particle–particle interactions}, which mediate communication without direct contact. As such, auto-chemotaxis provides a unified framework to study how memory and interaction-induced coordination jointly shape collective search.

Beyond its theoretical simplicity, auto-chemotaxis is biologically well motivated. Chemical signaling is one of the most elementary communication mechanisms available to microorganisms and underlies processes such as bacterial navigation \cite{berg,fu2018nc}, immune cell migration \cite{LeukocyteDynamics,insall2022ticb}, and collective cell dynamics during development \cite{TissueMigration}. 
Chemotactic interactions have also been realized in synthetic matter, like self-propelled microdroplets that communicate via chemorepulsive trails \cite{hokmabad2022chemotactic} or colloidal particles that leave phase-change trails \cite{Tunable}, mimicking tunable pheromone interactions as among ants. Synthetic chemotaxis in active matter systems generically lead to pattern formation like dynamic clusters and waves \cite{liebchen1,liebchen2,liebchen3} and has recently been studied intensively 
\cite{kranz2019trail, hokmabad2022chemotactic, mokhtari2022spontaneous, barbier2022self, moreno2022single, meyer2023alignment}. Moreover, it is a natural example of non-reciprocal interactions, i.e. interactions that violate the {\it actio$=$reactio} principle and that became recently a major research focus 
\cite{nri-loewen,nri-marchetti,nri-golestanian,fruchart2021non,nri-rao,nri-kreienkamp,nri-matthieu,nri-jiwon,nri-atul,nri-jaedong,nri-chulung}.

From this perspective, auto-chemotactic repulsion represents a minimal but ubiquitous strategy by which searchers can share information about explored regions and suppress redundant exploration.
Using this framework, we show that collective search efficiency is controlled by a nontrivial balance between memory, directional persistence, and spatial organization. We identify two qualitatively distinct regimes. In a weak-memory regime, efficient search emerges from interaction-induced spatial ordering that effectively partitions the search domain. In a strong-memory regime, long-lived but decaying chemical trails suppress revisits through collective self-avoidance, even in the absence of global order.
In both cases, optimal search does not coincide with maximal memory or maximal spatial order. Instead, it arises from an intermediate regime in which memory and interactions are strong enough to reduce redundancy, yet sufficiently weak to avoid over-constraining motion.

All results are obtained numerically. Despite the simplicity of the microscopic rules, collective auto-chemotactic search is a history-dependent many-body problem with non-Markovian dynamics, which makes an analytical treatment intractable.

The paper is organized as follows. In Sec.~\ref{sec:models} we introduce two microscopic models of auto-chemotaxis. Section~\ref{sec:observables} defines the key observables used throughout the paper. In Sec.~\ref{sec:regimes} we describe how to distinguish between weak and strong memory regimes. Section~\ref{sec:low_memory} presents results in the weak-memory regime, where the dynamics is dominated by effective long-range repulsive interactions and self-propulsion. In Sec.~\ref{sec:strong_memory} we investigate the strong-memory regime, in which chemical trails persist over long times, and highlight the resulting differences in search strategies. Finally, Sec.~\ref{sec:conclusion} summarizes our findings and discusses their broader implications.

\section{Models, Observables \& Memory Scales}

\subsection{Models of auto-chemotaxis}
\label{sec:models}

Auto-chemotactic motion couples chemical production along particle trajectories with orientation response to chemical gradients. To describe search by such agents, we employ both an off-lattice model, which captures realistic active Brownian dynamics, and a lattice model, which can access much lower densities and strong memory regimes.

\subsubsection{Off-lattice model}

We consider a two-dimensional off-lattice active Brownian particle (ABP) model in which each self-propelled disk secretes a diffusing chemical field that mediates auto-chemotactic repulsion \cite{taktikos2011pre,taktikos2012pre}. Particle $i$ with position $\mathbf{r}_i$ moves with constant propulsion speed $v_0$ along an intrinsic orientation $\mathbf e_i(t)=\left[\cos\varphi_i(t),\,\sin\varphi_i(t)\right]^{\top}$, undergoing rotational diffusion. Its translational dynamics obey
\begin{equation}
    \label{eq:ABP1}
    \dot{\bf r}_i(t)=v_0\,{\bf e}_i(t)  
    +\sum_{{j=1}\atop{j\neq i}}^{N} \frac{{\bf f}_{ij}}{\gamma_t}\,,
\end{equation}
where ${\bf f}_{ij}$ is a short-range repulsive force preventing overlap.

Each particle deposits chemical at its instantaneous location. The field $c(\mathbf{r},t)$ diffuses with coefficient $D_c$ and decays with rate $\alpha_c$:
\begin{align}
    \label{eq:chem_diffusion}
    \dot{c} =& D_c\nabla^2  c  - \alpha_c c + h_c \sum_{i=1}^{N}\delta(\mathbf{r}-\mathbf{r}_i(t))\,.
\end{align}
The particles respond to gradients of their own collective field through a chemotactic torque,
\begin{equation}
    \label{eq:chemotaxis}
    \dot{\varphi}_i(t)=\beta \,
    [\,\nabla c({\bf r}_i(t),t) \times {\bf e}_i(t)\,]_z 
    + \sqrt{2D_r}\xi_i(t)\,,
\end{equation}
which rotates the particle orientation away from chemical gradients for $\beta>0$ (chemo-repulsion). Here, $\varphi_i(t)$ denotes the polar angle of the propulsion direction. The second term on the r.h.s. is the usual rotational noise for ABPs. We simulate $N$ particles in a square box of length $L$ using periodic boundary conditions. Unless otherwise stated, lengths are measured in units of the particle radius $a$ and time in units of the rotational diffusion time $1/D_r$. The packing fraction is defined as $\phi = \pi a^2 N/L^2$. In the appendix \ref{app:models} we provide additional details of the ABP implementation.

For an ideal ABP, the \emph{bare persistence length} is defined as $l_p=v_0/(aD_r)$. Interactions may modify the orientational dynamics and, in principle, also the propulsion speed. Thus, we define an effective persistence length as
\begin{equation}
l_p^{\mathrm{eff}} = \frac{v_0^{\mathrm{eff}}}{a D_r^{\mathrm{eff}}}\,,
\end{equation}
where $v_0^{\mathrm{eff}}$ and $D_r^{\mathrm{eff}}$ denote the effective propulsion speed and rotational diffusion coefficient, respectively. The latter is obtained from the decay of the orientation autocorrelation function, $\langle\mathbf e(0)\cdot\mathbf e(t)\rangle\propto\exp(-D_r^{\mathrm{eff}}t)$. In our auto-chemotactic model, the dominant effect of the trails is a modification of the orientational dynamics. Direct repulsive interactions can additionally affect the propulsion speed, particularly in dense or clustered states. However, the present work focuses on dilute and spatially homogeneous regimes, where persistent blocking of particles is absent and $v_0^{\mathrm{eff}}\simeq v_0$. Consequently, variations of $l_p^{\mathrm{eff}}$ are primarily controlled by changes in $D_r^{\mathrm{eff}}$.

This continuous model connects directly to the standard ABP model and, in the weak-memory regime, reduces effectively to an ABP with Yukawa-type repulsive interactions.

\subsubsection{Lattice model}

As a complementary approach to the continuous off-lattice model, we use a discrete auto-chemotactic random walk~\cite{meyer2023alignment}. We place $N$ walkers on the sites of a two-dimensional square lattice of size $L\times L$ with periodic boundary conditions. The state of a walker is described by the probability $\rho(i,t)$ of being at site $i$ at time $t$. Its dynamics are governed by the master equation
\begin{equation}
    \rho(i,t+\Delta t) = \sum_{j\in\mathcal{N}_i} p_{j\to i}(t)\,\rho(j,t),
\end{equation}
where $\mathcal{N}_i$ is the set of nearest neighbors of site $i$, and the discrete time step is set to $\Delta t=1$. Thus lengths are in units of the lattice spacing and time in units of $\Delta t$.

The transition probability from site $j$ to a neighboring site $i$ reads
\begin{equation}
    \label{eq:prob_lattice}
    p_{j\to i}(t) = \frac{\left(1+b_i\right)\,e^{-\beta c(i,t)}}{\sum_{k\in\mathcal{N}_j} \left(1+b_k\right)\,e^{-\beta c(k,t)}},
\end{equation}
where $c(i,t)$ is the chemical concentration at site $i$ and time $t$, and $\beta$ is the chemotactic coupling strength. For $\beta>0$ the walker preferentially moves towards sites with lower concentration thus generating chemo-repulsion, which is the case considered here. While no hard-core exclusion is implemented in the lattice model, repulsive chemotaxis dynamically suppresses multiple occupancy. We define the occupation fraction as $\varphi = N/L^2$. When comparing both models, we identify $\varphi$ with the off-lattice packing fraction $\phi$.

The concentration field evolves according to the discrete analogue of Eq.~(\ref{eq:chem_diffusion}) with diffusion constant $D_c$ and decay rate $\alpha_c$. At each time step, walkers deposit chemical on their current site. For $D_c=0$ and $\alpha_c=0$, deposited chemical persists indefinitely and the model reduces to a self-avoiding walk \cite{amit1983asymptotic}. In this limit, the variance of the chemical field grows linearly in time, a signature commonly associated with aging effects \cite{grassberger2017prl}. However, within the parameter range and time windows investigated here, we did not observe a systematic dependence of the mean first-passage time on the equilibration time.

The parameter \(b_i\) encodes an \emph{intrinsic} directional persistence: if the jump \(j \to i\) continues in the same direction as the previous step, then \(b_i = b\) (a constant), otherwise \(b_i = 0\) \cite{tejedor2012optimizing}. In the absence of chemotaxis (\(\beta = 0\)), the mean number of consecutive steps in the same direction defines the \emph{bare persistence length}
$l_p =(4+b)/3$. Chemotactic interactions modify this persistence: in case of chemo-repulsion (\(\beta \neq 0\)) walkers tend to maintain their direction for longer, leading to an \emph{effective persistence length} \(l_p^{\text{eff}}> l_p\).

The lattice and off-lattice models differ substantially in their microscopic dynamics. The purpose of considering both is not to establish quantitative equivalence, but to test the robustness of trail-mediated self-repulsion as a search mechanism. Agreement between the models should therefore be interpreted at a qualitative level.

\subsection{Core observables}
\label{sec:observables}

\subsubsection{Collective speedup}
\label{sec:def_speedup}

To quantify the benefit of collective search, we compare the mean first-passage time (MFPT) of $N$ interacting agents with that of $N$ non-interacting agents that couple only to their self-generated chemical field. We define the \emph{collective speedup} as
\begin{equation}
\gamma_N =  1 -\frac{\bar{T}_N}{\bar{T}^\text{ind}_N},
\end{equation}
quantifying the relative gain in performance due to interactions. A value of $\gamma_N > 0$ indicates that interaction between agents leads to faster search than a purely individualistic strategy. Here, $\bar{T}_N$ is the MFPT for the first of $N$ interacting agents to reach a randomly located target. The reference value $\bar{T}^\text{ind}_N$ is the MFPT for $N$ independent searchers, which can be obtained from single-agent search:
Let $S_1(t)$ denote the survival probability of the target in a single-agent search, i.e., the probability that the target has not been found by time $t$. For $N$ independent agents searching in parallel, the joint survival probability is $S_N(t) = [S_1(t)]^N$, and the corresponding MFPT becomes
\begin{equation}
\label{eq:TN_ind}
    \bar{T}_{N}^{\mathrm{ind}} = \int_{0}^\infty [S_1(t)]^N\, \mathrm{d}t\,.
\end{equation}
In the special case where $S_1(t)$ decays exponentially, this reduces to $\bar{T}_N^{\mathrm{ind}} = \bar{T}_1 / N$. However, this simple scaling can break down for large $N$ due to extreme-value effects associated with the closest searcher \cite{ro2017pre,ro2022pre}.

\subsubsection{Spatial Order Parameter}
\label{sec:def_spatial_order}

Since the spatial arrangement of searchers affects collective search performance, we define a quantitative measure of spatial organization that distinguishes homogeneous from inhomogeneous particle distributions~\cite{monchaux2010pof}. Given an instantaneous configuration of $N$ searchers with positions $\left\{ \mathbf{r}_i \right\}$, we compute the Voronoi tessellation of the full domain of size $L^2$. Each searcher $\mathbf{r}_i$ is associated with a Voronoi cell of area $A_i$, containing all points closer to $\mathbf{r}_i$ than to any other particle. We normalize the cell areas by defining $\mathcal{A}_i = A_i / \langle A \rangle$, where the mean area is $\langle A \rangle = L^2/N$, and compute the standard deviation of the normalized areas as
\begin{equation}
\sigma_\mathcal{A} = \sqrt{ \langle \mathcal{A}^2 \rangle - \langle \mathcal{A} \rangle^2 }\,.
\end{equation}

For randomly distributed particles (Poisson process), this value is known from numerical studies to be $\sigma_\mathcal{A}^{\mathrm{ind}} = \sqrt{0.28} \approx 0.53$ in the limit of large $N$~\cite{ferenc2007phys_a}. We define \emph{spatial order parameter} as
\begin{equation}
    \eta_\mathcal{A} = 1 - \frac{\sigma_\mathcal{A}}{\sigma_\mathcal{A}^{\mathrm{ind}}}\,.
\end{equation}
This quantity measures deviations from random spatial structure: $\eta_\mathcal{A} > 0$ indicates a more regular (uniform) spacing than in the random case, while $\eta_\mathcal{A} < 0$ indicates clustering. In the present work, we focus exclusively on parameter regimes with $\eta_{\mathcal A}\ge0$. For example, if particles are uniformly assigned to $N$ distinct subdomains as in a split search process, but randomly distributed within each subdomain, the spatial order is found to be $\eta_\mathcal{A}^{\mathrm{split}} \approx 0.53$.

%NEW
\subsection{Memory regimes}
\label{sec:regimes}

\begin{figure*}
    \centering
    \begin{minipage}{0.45\linewidth}
        \includegraphics[width=\linewidth]{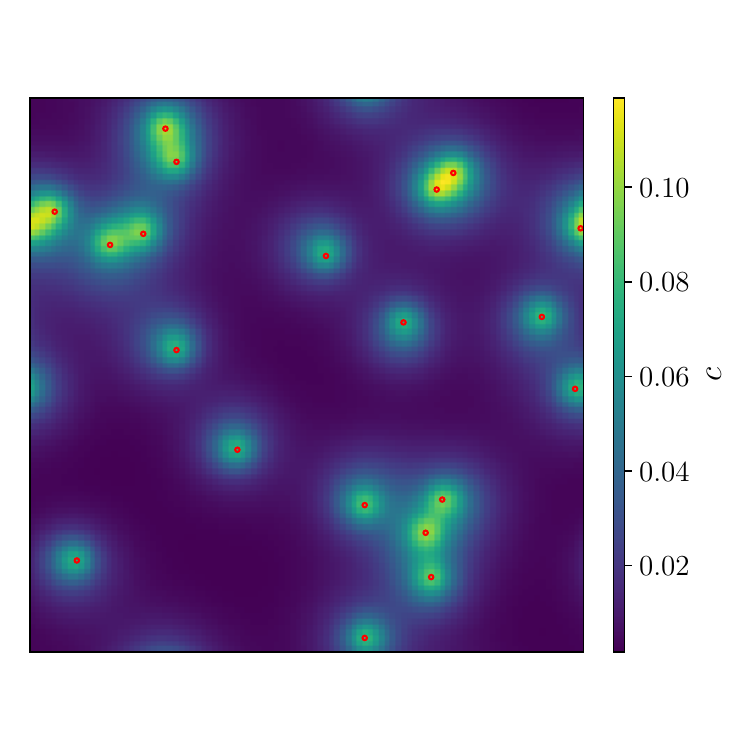}
    \end{minipage}
    \begin{minipage}{0.45\linewidth}
    \includegraphics[width=\linewidth]{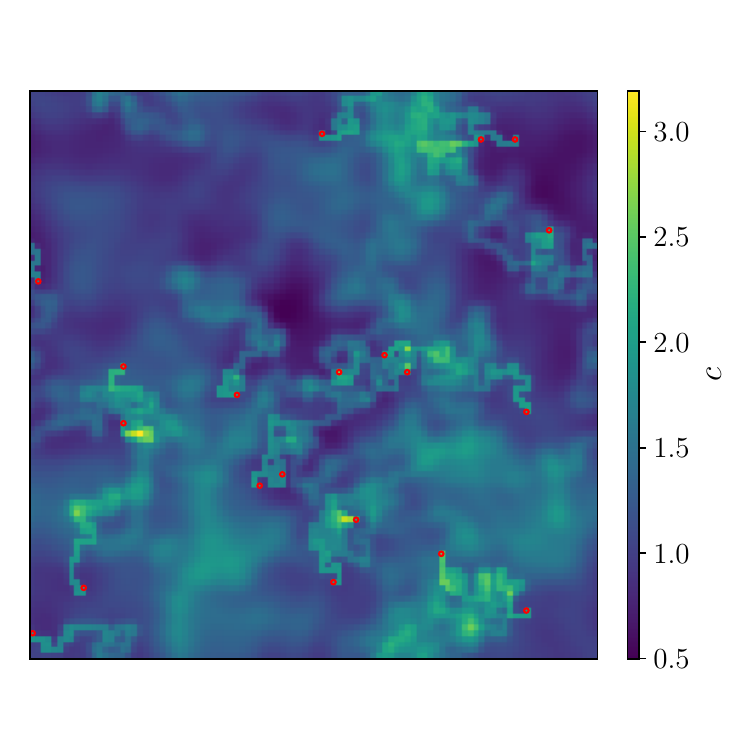}
    \end{minipage}
    \caption{Simulation snapshots of the lattice auto-chemotactic model illustrating the weak- and strong-memory regimes. (left) Weak-memory regime ($D_c=2.5$, $\alpha_c=0.1$, $2\sqrt{D_c\alpha_c}=1$), where the chemical field is short-lived and does not form persistent trails. (right) Strong-memory regime ($D_c=0.01$, $\alpha_c=0.001$, $2\sqrt{D_c\alpha_c}=0.0063$), where persistent, asymmetric trails form along past trajectories and overlap with those of other walkers. In both panels, $N=20$ walkers move in a box of size $L=100$ ($\varphi=0.002$) with coupling strength $\beta=10$. The background color encodes the chemical concentration field and the red circles indicate walker positions.}

\label{fig:c_snapshots}
\end{figure*}

The nature of the interaction between auto-chemotactic searchers is set by two characteristic lengths: a \emph{chemical decay length}  
\begin{equation}
\lambda_{\mathrm{decay}} = \sqrt{\frac{D_c}{\alpha_c}}\,,
\label{eq:decay_length}
\end{equation}
the typical distance a chemical molecule diffuses before decaying, and an \emph{advective length}
\begin{equation}
\lambda_{\mathrm{adv}} = \frac{2D_c}{v_0}\,,
\label{eq:adv_length}
\end{equation}
the distance over which diffusion redistributes the chemical before the source moves away.

To quantify the resulting asymmetry of the chemical profile, consider a walker moving at constant velocity $v_0$. In its comoving frame, the steady-state concentration decays asymmetrically ahead and behind (see Appendix~\ref{app:concentration_profile}). The ratio of the two length scales controls the fore–aft asymmetry of the profile. In the \emph{zero-memory regime}, if
\[
\lambda_{\mathrm{decay}} \ll \lambda_{\mathrm{adv}}
\quad \text{or} \quad
2\sqrt{D_c \alpha_c} \gg v_0\,,
\]
the chemical field is short-lived and effectively isotropic around each walker. Interactions are effectively instantaneous and correspond to a Yukawa-type potential in two dimensions. For intermediate values $\lambda_{\mathrm{decay}} \simeq \lambda_{\mathrm{adv}}$ or $2\sqrt{D_c \alpha_c} \simeq v_0$, the system enters a weak-memory regime in which the field retains a finite fore–aft asymmetry but remains short-lived in time, so that the particle primarily interacts with its recently emitted field rather than with a persistent trail. In contrast, in the \emph{strong-memory regime}, if
\[
\lambda_{\mathrm{decay}} \gg \lambda_{\mathrm{adv}}
\quad \text{or} \quad
2\sqrt{D_c \alpha_c} \ll v_0\,,
\]
the field becomes strongly asymmetric, forming long-lived trails along past trajectories. Walkers then interact with their own history, leading to trail-mediated self-avoidance and dynamically structured environments. For the lattice model, where $v_0=1$ by construction, the criterion reduces to $2\sqrt{D_c \alpha_c} \gg 1$ (zero memory) and $2\sqrt{D_c \alpha_c} \ll 1$ (strong memory). Figure~\ref{fig:c_snapshots} shows representative configurations of the chemical field in both regimes. The qualitative contrast between localized profiles and persistent trails anticipates the distinct search mechanisms analyzed below.

\section{Weak memory regime}
\label{sec:low_memory}

\subsection{Single searcher}

\begin{figure}
    \centering
    \includegraphics[width=.9\linewidth]{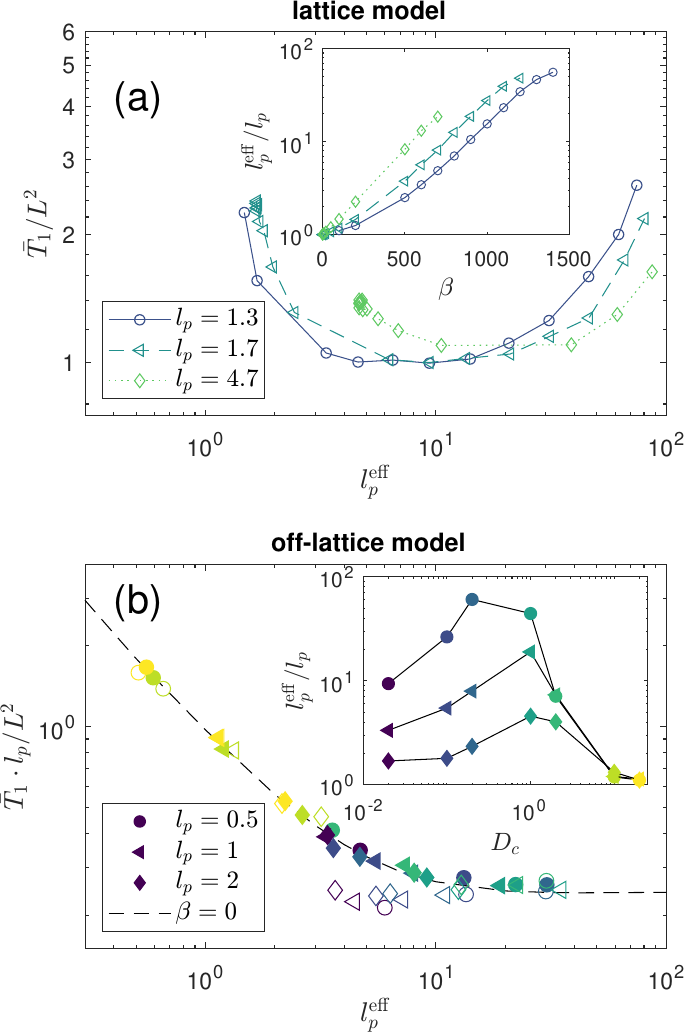}
    \caption{Single-agent search as a function of effective persistence length $l_p^\mathrm{eff}$. Mean first-passage time (MFPT) $\bar{T}_1$ versus $l_p^\mathrm{eff}$ in (a) the lattice model, where $l_p^\mathrm{eff}$ is tuned by the chemotactic coupling strength $\beta$. Colors correspond to different values of the bare persistence length $l_p$. (b) Off-lattice model, where $l_p^\mathrm{eff}$ is tuned by the chemical diffusivity $D_c$. Colors indicate different values of $D_c$. Insets show how $l_p^\mathrm{eff}$ depends on the respective control parameter. For finite chemical decay, the data collapse onto a common master curve consistent with ABP behavior. Empty symbols in (b) correspond to $\alpha_c=0$, where strong-memory effects lead to systematic deviations from the master curve. In all cases with finite chemical decay, the parameters lie in the finite-memory regime defined in Sec.~\ref{sec:regimes}, where no persistent trail accumulation occurs.}
    \label{fig:single_searcher}
\end{figure}

In the weak–memory regime, $\lambda_{\mathrm{decay}} \simeq \lambda_{\mathrm{adv}}$, which lies between the zero-memory and the strong-memory regimes, the chemical field exhibits a finite fore–aft asymmetry but remains short-lived. The particle interacts primarily with the local gradient of its recently emitted field rather than with a long-lived record of its past trajectory. Self-interaction, therefore, produces a change in persistence but does not introduce genuine long-term memory. We first consider the single-particle case ($N=1$) to isolate self-interaction effects.

Several microscopic parameters influence the effective persistence length $l_p^\mathrm{eff}$ by affecting gradient strength or shape. In each model we vary only one control parameter ($\beta$ in the lattice model, $D_c$ in the off-lattice model), noting that the results depend only on the resulting persistence length, not on the specific tuning mechanism.

In the lattice model, increasing the chemotactic coupling strength $\beta$ enhances the response to the fore-aft asymmetry, biasing the motion to straighter trajectories and thereby increasing $l_p^\mathrm{eff}$ beyond its bare value $l_p$, see inset in Fig.~\ref{fig:single_searcher}(a). In the limit of very strong coupling, the particle approaches a ballistic regime.

In the off-lattice model $\beta$ is fixed, and instead we vary the chemical diffusion constant $D_c$. Changing $D_c$ changes the shape of the self-generated field, which in turn modifies the effective rotational diffusion constant $D_r^\mathrm{eff}$ and thereby $l_p^\mathrm{eff} = v_0 / (a D_r^\mathrm{eff})$. For small $D_c$, the chemical signal remains highly localized and the particle rarely overlaps with its own trail; for large $D_c$ the signal becomes broad and shallow. In both limits, $D_r^\mathrm{eff} \to D_r$, and thus $l_p^\mathrm{eff} \to l_p$. As a result, $D_r^\mathrm{eff}$ exhibits a minimum, and correspondingly $l_p^\mathrm{eff}$ a maximum, at intermediate $D_c$, see inset of Fig.~\ref{fig:single_searcher}(b).

This change in persistence directly impacts search efficiency. When the mean first-passage time $\bar{T}_1$ is plotted against $l_p^\mathrm{eff}$, see Fig.~\ref{fig:single_searcher}(a,b), the data collapse onto a single master curve in both models, matching the behavior of a persistent random walker and a simple active Brownian particle (ABP). In the lattice model, $\bar{T}_1$ diverges as $l_p^\mathrm{eff} \to \infty$, consistent with persistent random walks~\cite{tejedor2012optimizing}. Such a divergence does not occur for finite persistence in the off-lattice model, where orientations are continuous.

These results demonstrate that, in the weak-memory regime, auto-chemotactic self-interaction acts primarily as a renormalization of persistence: search efficiency is controlled by $l_p^\mathrm{eff}$, independent of microscopic interaction details. However, in the absence of chemical decay ($\alpha_c = 0$), the equivalence with a simple ABP breaks down, see empty symbols in Fig.~\ref{fig:single_searcher}(b). The accumulating concentration field introduces strong long-term memory, and systematic deviations from the ABP master curve emerge. In particular, particles with small $D_c$ can attain MFPTs below the ABP optimum and approach the ballistic limit $\bar{T}_1 = L^2/(4 l_p)$. This shows that in the strong-memory regime, the self-generated trail does more than renormalize persistence but qualitatively modifies the search strategy. The collective consequences of this long-memory behavior are discussed in Sec.~\ref{sec:strong_memory}.

% NEW
\subsection{Collective search}

Collective search performance emerges from the interplay between two competing ingredients: the tendency of agents to spread across space and avoid redundant exploration, and their ability to move persistently to efficiently scan unexplored regions. In auto-chemotactic systems, repulsive interactions promote spatial separation, while self-interaction with chemical trails enhances directional persistence. These mechanisms are intrinsically coupled, so that collective efficiency cannot be optimized by tuning either persistence or spatial order in isolation.

\begin{figure}
    \centering
    \includegraphics[width=.9\linewidth]{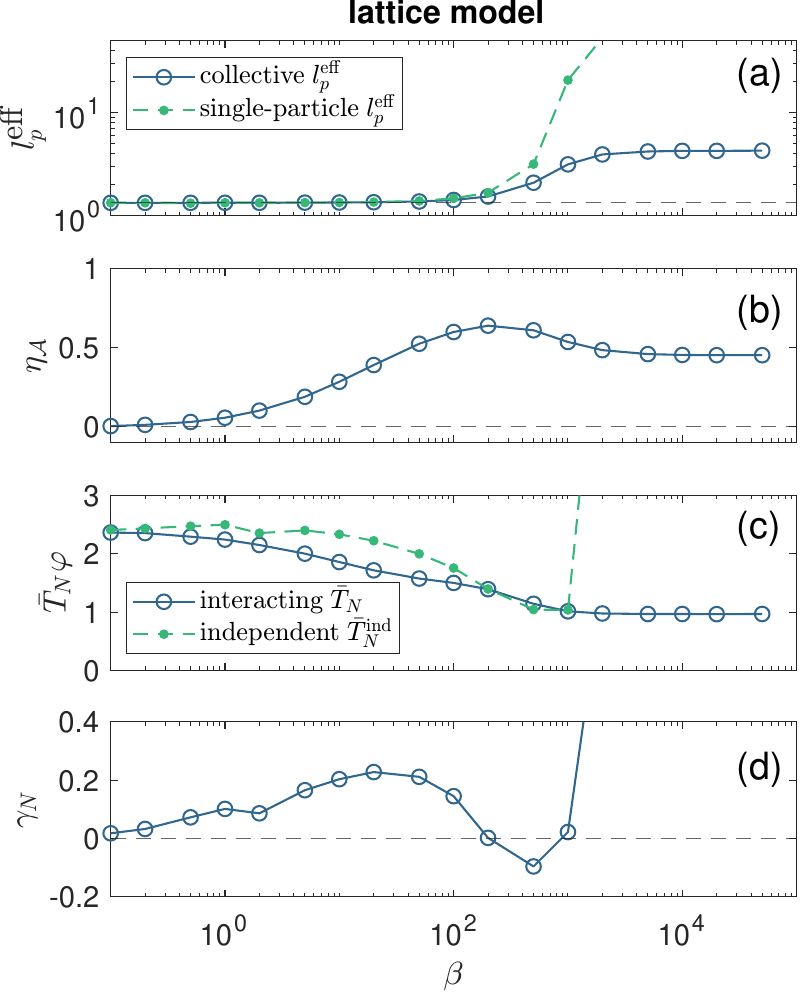}
    \caption{(a) Effective persistence length $l_p^{\mathrm{eff}}$ of interacting auto-chemotactic lattice random walkers (circles) compared to the single-particle value (dots) as a function of chemotactic coupling $\beta$. Collisions suppress the growth of $l_p^{\mathrm{eff}}$ at strong coupling. (b) Spatial order parameter $\eta_\mathcal{A}$. (c) Collective MFPT $\bar{T}_N$ (circles) and the independent-search reference $\bar{T}_N^{\mathrm{ind}}$ (dots). (d) Cooperative speedup $\gamma_N$. Parameters: bare persistence $l_p=4/3$ and searcher occupation fraction $\varphi=0.012$.}
    \label{fig:lpeff_order_mfpt_speedup_vs_beta}
\end{figure}

We first vary the chemotactic coupling $\beta$ at fixed density to probe how this balance shapes collective dynamics in the lattice model, see Fig.~\ref{fig:lpeff_order_mfpt_speedup_vs_beta}. At weak coupling, increasing $\beta$ primarily induces spatial order while leaving the effective persistence length $l_p^{\mathrm{eff}}$ unchanged [Fig.~\ref{fig:lpeff_order_mfpt_speedup_vs_beta}(a,b)].  A constant $l_p^{\mathrm{eff}}$ indicates that the initial performance gain is driven by spatial separation rather than persistence enhancement. Repulsion reduces redundant exploration, leading to a decrease of the MFPT [Fig.~\ref{fig:lpeff_order_mfpt_speedup_vs_beta}(c)] and a corresponding increase of the cooperative speedup $\gamma_N$ [Fig.~\ref{fig:lpeff_order_mfpt_speedup_vs_beta}(d)]. At intermediate coupling, self-interaction enhances directional persistence. In contrast to the single-particle case, however, inter-particle collisions bound the growth of $l_p^{\mathrm{eff}}$, which saturates at a finite value $l_p^\infty \equiv \lim_{\beta\to\infty} l_p^{\mathrm{eff}}(\beta,\varphi)$ [Fig.~\ref{fig:lpeff_order_mfpt_speedup_vs_beta}(a)]. In this regime, the single-search MFPT is already close to its persistence-controlled optimum, and $\bar{T}_N \approx \bar{T}_N^{\mathrm{ind}}$. Collective interactions therefore provide little additional advantage, and the speedup $\gamma_N$ decreases and can even become negative. At strong coupling, a distinct lattice-specific effect appears. As $\beta\to\infty$, single-particle motion becomes ballistic and both $\bar{T}_1$ and $\bar{T}_N^{\mathrm{ind}}$ diverge, see [Fig.~\ref{fig:single_searcher}(a)]. In contrast, collisions regularize persistence in the interacting system, allowing particles to explore new regions so that the collective MFPT $\bar{T}_N$ remains finite. This leads again to an increase of $\gamma_N$ at large $\beta$, even though neither spatial order nor persistence grow further.

The $\beta$-scan demonstrates that spatial order and effective persistence are intrinsically coupled: large values of both cannot be achieved simultaneously in the weak-memory regime (see Appendix~\ref{app:order_persistence}). Because efficient search requires both persistent motion and suppression of redundant exploration, the optimal strategy corresponds to a specific balance between them. Each pair of control parameters $(l_p,\beta)$ therefore selects a distinct point in $(\eta_\mathcal{A}, l_p^{\mathrm{eff}})$ space. We now determine, for each density, the combination that minimizes the MFPT.

\begin{figure}
    \centering
    \includegraphics[width=.8\linewidth]{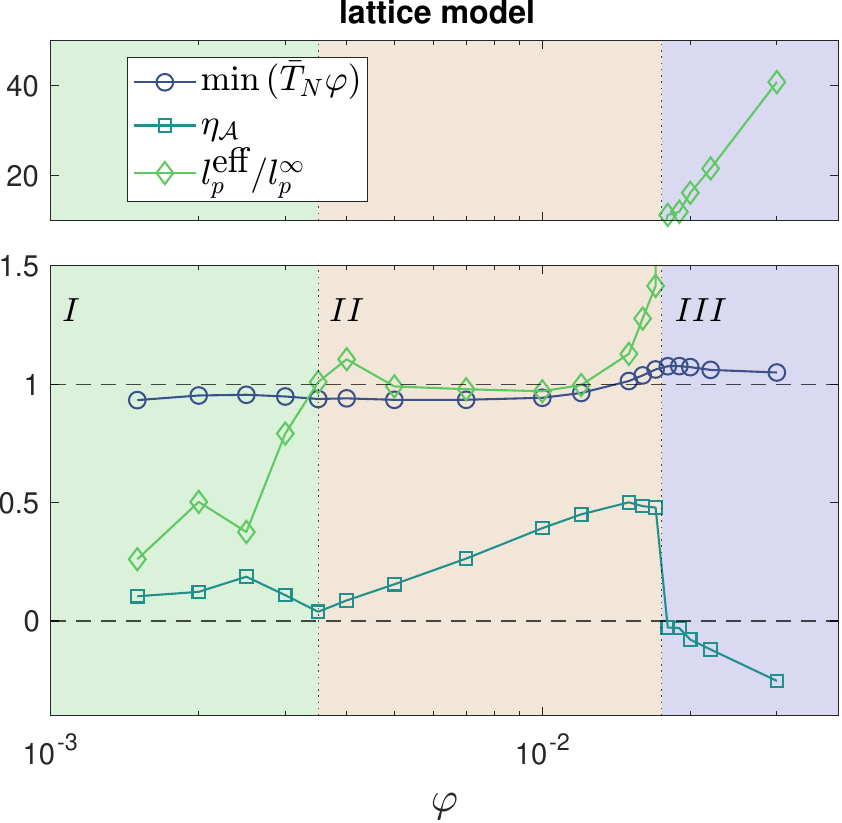}
    \caption{Minimum MFPT $\min_{\beta,l_p}(\bar{T}_N\varphi)$ as a function of density $\varphi$ in the lattice model. The corresponding optimal $l_p^\mathrm{eff}$ and $\eta_\mathcal{A}$ are shown. Here $l_p^\infty$ denotes the effective persistence length in the strong-coupling limit $\beta\to\infty$. Shaded regions indicate distinct density regimes with different optimal search strategies.}
    \label{fig:optimal_mfpt_vs_density}
\end{figure}

Figure~\ref{fig:optimal_mfpt_vs_density} shows the minimum MFPT $\min_{\beta,l_p}(\bar{T}_N\varphi)$ as a function of density $\varphi$ in the lattice model, together with the corresponding optimal values of $l_p^{\mathrm{eff}}$ and $\eta_\mathcal{A}$. Three density regimes emerge, each characterized by a distinct optimal compromise between persistence and spatial separation. At low density, optimal search occurs at moderate coupling: repulsion generates finite spatial order while persistence remains sufficiently large to ensure efficient exploration. At intermediate densities, strong coupling becomes optimal. Particles move nearly ballistically between collisions, spatial order approaches its maximal attainable value, and the effective persistence saturates at $l_p^\infty$. At high density, however, strong coupling is detrimental. Frequent collisions suppress persistence, and the optimal strategy shifts toward weak coupling, $\beta \approx 0$, where interactions are minimal and persistence is controlled by the bare value $l_p$. Optimal search is achieved at an intermediate balance between persistence and spatial separation, rather than at maximal spatial order.

\begin{figure}
    \centering
    \includegraphics[width=1\linewidth]{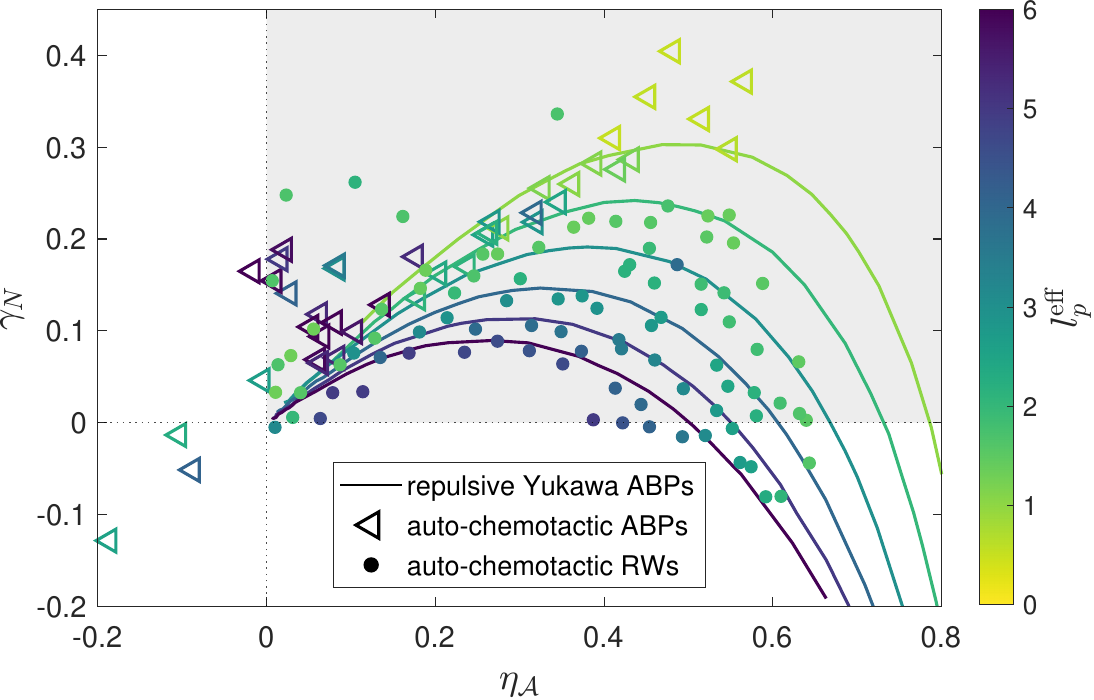}
    \caption{Cooperative search speedup $\gamma_N$ versus spatial order $\eta_\mathcal{A}$ for repulsive Yukawa active Brownian particles (ABPs), auto-chemotactic ABPs, and auto-chemotactic random walkers (RWs). Color indicates the effective persistence length $l_p^\mathrm{eff}$. The gray region corresponds to $\gamma_N>0$ and $\eta_\mathcal{A}>0$, where cooperative search is enhanced and spatial order exceeds that of independent searchers. Colored curves show the Yukawa ABP reference data for different values of $l_p^\mathrm{eff}$. While the models do not collapse onto a universal curve, they exhibit a common qualitative trend: cooperative speedup is typically largest at intermediate spatial order and decreases again at high order. The lattice occupation fraction is $\varphi=0.012$ and the off-lattice packing fraction is $\phi=0.01$.}
    \label{fig:order_speedup_all}
\end{figure}

While the collective MFPT $\bar{T}_N$ quantifies absolute group performance, the cooperative speedup $\gamma_N$ directly measures the efficiency gain due to interactions relative to independent searches. To test whether the relation between spatial order and cooperative gain depends on the microscopic interaction mechanism, we compare three distinct models: repulsive Yukawa ABPs, auto-chemotactic ABPs, and the lattice auto-chemotactic random-walker model. Figure~\ref{fig:order_speedup_all} shows $\gamma_N$ as a function of spatial order $\eta_\mathcal{A}$. Although the microscopic dynamics differ substantially, all three models exhibit the same qualitative trend. A moderate degree of spatial order is generally associated with enhanced cooperative speedup, indicating that ordering suppresses redundant exploration and promotes a more efficient division of search effort. However, this trend is not monotonic. When the spatial order becomes too strong, particle motion becomes increasingly constrained and the cooperative speedup can saturate or even decrease due to caging and over-constrained motion. The decrease of $\gamma_N$ at large $\eta_\mathcal{A}$ occurs entirely within the homogeneous regime and is therefore not associated with clustering or band formation. The agreement between the models is qualitative rather than quantitative. While the data do not collapse onto a single universal curve, most auto-chemotactic data lie close to the reference curves of repulsive Yukawa ABPs. For a given $l_p^{\mathrm{eff}}$, only a finite range of speedups is attainable, and optimal performance occurs at intermediate spatial order rather than maximal ordering, consistent with the density-dependent optimization shown in Fig.~\ref{fig:optimal_mfpt_vs_density}.

\begin{figure}
    \centering
    \includegraphics[width=.8\linewidth]{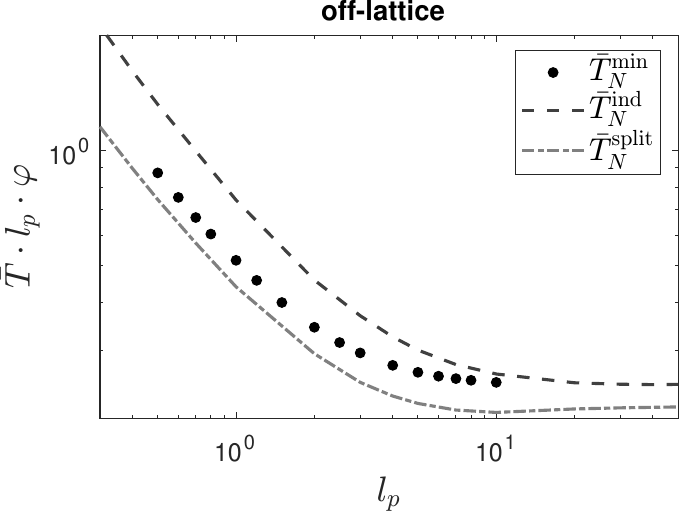}
    \caption{Mean first-passage (MFPT) time for repulsive active Brownian particles (ABPs). Minimum MFPT $\bar{T}_N^{\mathrm{min}}$, obtained by optimizing over the strength of the repulsive Yukawa interaction, as a function of the bare persistence length $l_p$ (dots). For comparison, the dashed line shows the MFPT of independent ABPs $\bar{T}_N^{\mathrm{ind}}$, while the dot-dashed line corresponds to MFPT $\bar{T}_N^{\mathrm{split}}$ of an artificial domain-splitting strategy in which walkers explore non-overlapping subdomains. The packing fraction is $\phi=0.01$.}
    \label{fig:mfpt_vs_lp_yukawa}
\end{figure}

An even stronger test of mechanism-independence is obtained by comparing soft repulsive interactions with an extreme limit in which spatial separation is enforced geometrically. Figure~\ref{fig:mfpt_vs_lp_yukawa} shows the minimum MFPT for Yukawa-interacting active Brownian particles, optimized over interaction strength, together with an artificial domain-splitting strategy in which each walker explores a non-overlapping subdomain. In practice, this is implemented by comparing $N$ particles in a square box of side length $L$ with a single particle in a box of side length $L/\sqrt{N}$, such that the area per searcher is identical in both cases. Despite the absence of explicit interactions in the latter scenario, the resulting MFPT reductions are quantitatively similar. This demonstrates that collective search enhancement does not depend on the microscopic form of repulsion, but rather on its effective role in allocating exploration space and regulating persistence.

Taken together, these results show that collective search is optimized by tuning interactions to enhance persistence without enforcing maximal spatial order.

\section{Strong-memory regime}
\label{sec:strong_memory}

\begin{figure}
    \centering
    \includegraphics[width=\linewidth]{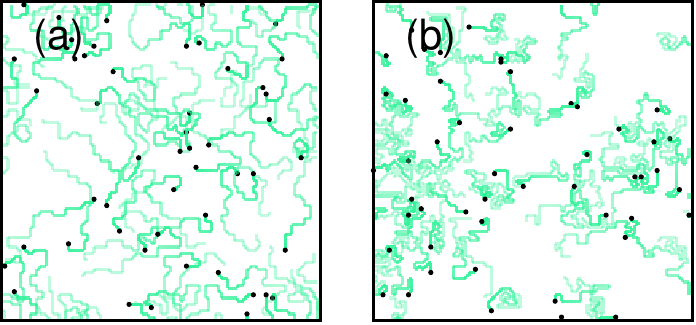}
    \caption{Representative trajectories of auto-chemotactic lattice random walkers of duration $T=60$ at density $\varphi=0.005$ and coupling strength $\beta=50$. (a) Weak-memory regime at $(D_c,\alpha_c)=(10^{-1},10^{-1})$, where chemical traces are short-lived and walkers do not effectively avoid previously explored regions, leading to frequent intersections between trajectories of different walkers. (b) Strong-memory regime at $(D_c,\alpha_c)=(10^{-3},10^{-3})$, where long-lived and localized chemical trails suppress crossings with both the walker’s own past trajectory and the trails produced by others, resulting in pronounced collective self-avoidance.}
    \label{fig:trajectories}
\end{figure}

A strong-memory regime arises when the chemical trail left by a walker is sufficiently sharp and asymmetric to encode spatially resolved information about its recent trajectory.
This occurs when the advective screening length is much smaller than the chemical decay length, $\lambda_{\mathrm{adv}} \ll \lambda_{\mathrm{decay}}$, see Eqs.~(\ref{eq:decay_length})–(\ref{eq:adv_length}).
In our lattice model with propulsion speed $v_0=1$, this criterion reduces to $2\sqrt{D_c\alpha_c} \ll 1$.
Except in limiting cases where diffusion is so large that chemical trails become spatially smeared, or decay is so fast that traces vanish almost instantly, this condition reliably predicts the onset of strong memory.
In this regime, walkers remain sensitive to chemical signals produced far in the past, and their dynamics is qualitatively distinct from that in the weak-memory limit. Representative trajectories of the two regimes are shown in Fig.~\ref{fig:trajectories}.

An operational signature of strong memory is provided by the survival probability of targets.
For sufficiently small $D_c$ and $\alpha_c$, the survival probability exhibits a compressed-exponential decay,
\begin{equation}
S_N(t) = \exp\!\left[-(t/T)^{\xi}\right],
\end{equation}
with exponent $\xi>1$, indicating a strong suppression of long search times. As $D_c$ or $\alpha_c$ increase, the trail becomes less informative, $\xi$ decreases toward unity, and the decay crosses over to a purely exponential form characteristic of weak memory. Across the explored parameter range, the condition $\xi>1$ coincides with $2\sqrt{D_c\alpha_c}\ll 1$, providing a consistent identification of the strong-memory regime, see Fig.~\ref{fig:MFPT_memory}(a) and details in Appendix~\ref{app:regimes}.

In the strong-memory regime of our lattice model, chemical trails become narrow and localized, which facilitates frequent reorientations and can reduce the effective persistence length. Nevertheless, long-lived trails continue to repel the walker from previously visited regions, so that self-avoidance and the suppression of revisits persist even in the absence of large directional persistence [Fig.~\ref{fig:trajectories}(b)].

A more microscopic characterization of strong memory is provided by the suppression of self-crossings \cite{arteca1999pre}.
Long-lived chemical trails effectively repel the walker from its own previously visited locations, strongly reducing the number of times its trajectory intersects its past path [Fig.~\ref{fig:trajectories}(b)].
To quantify this effect, we measure the average number of crossings $\bar{N}_{\mathrm{cross}}$ between a walker’s trajectory and previously deposited chemical trails during the time until the target was found, and compare it to the corresponding value $\bar{N}_{\mathrm{cross}}^{\mathrm{ind}}$ for a random walker with the same effective persistence length. We define the normalized crossing reduction
\begin{equation}
\chi = 1 - \frac{\bar{N}_{\mathrm{cross}}}{\bar{N}_{\mathrm{cross}}^{\mathrm{ind}}} ,
\end{equation}
so that $\chi>0$ indicates reduced self-crossing and enhanced self-avoidance. While the exponent $\xi$ characterizes the global survival statistics, the crossing reduction $\chi$ provides a local, trajectory-level measure of memory-induced self-avoidance.

Strong memory substantially enhances search efficiency, but remarkably, infinite memory does not. In the limit $D_c=\alpha_c=0$, corresponding to a true self-avoiding walk \cite{amit1983asymptotic}, chemical trails never decay and permanently exclude previously visited regions. 
While this completely suppresses revisits at short times, it ultimately becomes detrimental: closed loops of chemical traces form effective barriers that either trap the walker within its own trail network (self-caging) \cite{hokmabad2022chemotactic} or temporarily isolate unexplored regions from the accessible domain.
Both effects reduce the effective search space and increase the mean first-passage time (MFPT).
Efficient search therefore emerges at \emph{finite} memory strength: chemical traces must persist long enough to prevent frequent revisits, yet decay on a timescale that allows re-entry into regions that were previously enclosed.
This competition explains why the minimal MFPT lies inside the strong-memory regime but away from the self-avoiding limit [Fig.~\ref{fig:MFPT_memory}(a)], highlighting the importance of optimal forgetting. While the ideal space-covering lower bound $\bar{T}_1 = L^2/2$ cannot be surpassed, finite memory allows MFPT values that are much closer to this bound than those obtained with permanently self-avoiding trajectories [Fig.~\ref{fig:MFPT_memory}(b)].

\begin{figure*}
    \centering
    \includegraphics[width=\linewidth]{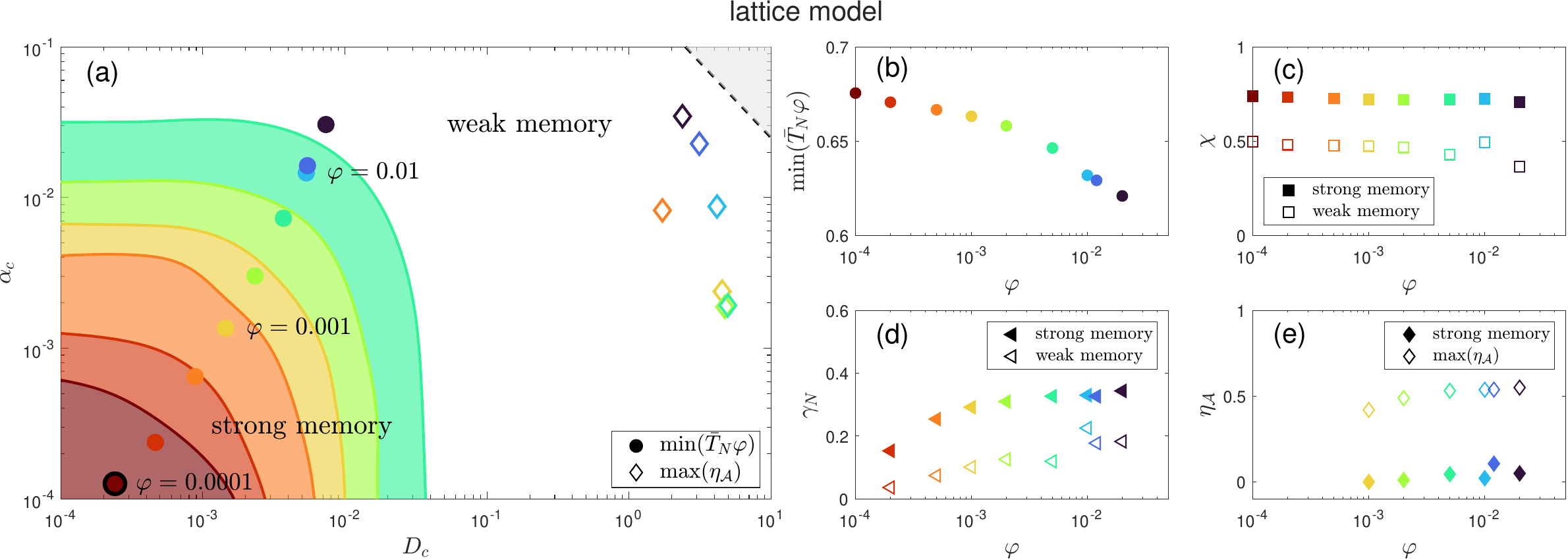}
    \caption{Strong-memory effects on collective search in the lattice model. (a) Location of the optimal mean first-passage time $\min(\bar{T}_N\varphi)$ (filled circles) and of the maximal spatial order $\max(\eta_\mathcal{A})$ (open diamonds) in the $(D_c,\alpha_c)$ plane for increasing searcher density $\varphi$. The upper-right shaded region corresponds to weak memory, where $2\sqrt{D_c\alpha_c}>1$ and chemotactic memory effects are negligible. The colored contours delimit the strong-memory regime ($\xi>1$); the enclosed region expands with increasing searcher density $\varphi$. Colors encode the searcher density $\varphi$ and are used consistently across all panels; representative densities are indicated in panel (a). The single-searcher case is indicated by a black outlined symbol and corresponds to $\varphi=10^{-4}$ in a system of size $L=100$. (b) Optimal search cost $\min(\bar{T}_N\varphi)$ as a function of density $\varphi$. (c) Path-crossing reduction parameter $\chi_{\mathrm{cross}}$, averaged over the strong- and weak-memory regimes. (d) Cooperative speedup $\gamma_N$, averaged over the strong- and weak-memory regimes. (e) Average spatial order $\eta_\mathcal{A}$ within the strong-memory regime compared to the global maximum of $\max(\eta_\mathcal{A})$. The chemotactic coupling strength is $\beta=50$ and the bare persistence length is $l_p=4/3$.}
    \label{fig:MFPT_memory}
\end{figure*}

To isolate the role of memory from chemotactic interactions, we also considered a minimal lattice model in which self-avoidance is imposed explicitly over a finite temporal window (a ``snake'' walk), see Appendix~\ref{app:snake} for details.
Figure~\ref{fig:snake_mfpt} shows the mean first-passage time as a function of the snake length $l_{\mathrm{snake}}$.
In all cases, the MFPT exhibits a clear minimum at finite memory.
When rescaled by system size, the optimal memory length collapses as
$l_{\mathrm{snake}}^\ast \sim L^2$, corresponding to the time required to explore the domain once.
This minimal model reproduces the same trade-off observed in auto-chemotactic search: insufficient memory leads to frequent revisits, whereas excessive memory causes self-caging and impedes exploration.

To further isolate the role of memory from a mere reduction of accessible search volume, we also investigated a minimal model in which lattice sites stochastically switch between accessible and inaccessible states, independently of the walker trajectory. While this introduces a fluctuating obstacle field that constrains motion, it does not systematically suppress redundant revisits, since site accessibility carries no information about the walker’s past exploration. Thus, this model fails to reproduce the strong reduction of the MFPT observed for snake walks or auto-chemotactic trails. This confirms that efficient search in the strong-memory regime relies on trajectory-dependent memory that actively penalizes revisits rather than on a passive reduction of the accessible search space.

\begin{figure}
\centering
\includegraphics[width=0.9\linewidth]{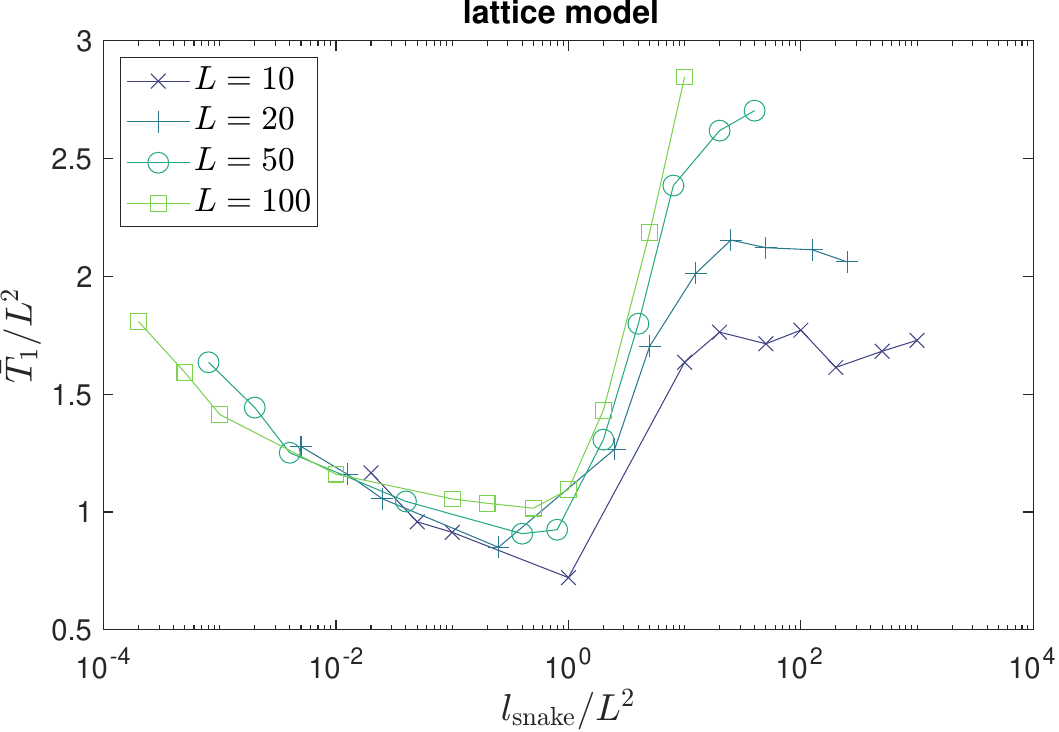}
\caption{Mean first-passage time for the snake \adam{lattice} model as a function of the normalized memory length $l_{\mathrm{snake}}/L^2$ for different system sizes $L$.
All curves exhibit a clear minimum at finite memory, with optimal values collapsing at $l_{\mathrm{snake}} \sim L^2$.
}
\label{fig:snake_mfpt}
\end{figure}

The same trail-mediated mechanism governs collective behavior. 
As the searcher density increases, the location of the minimal MFPT shifts toward larger values of $D_c$ and $\alpha_c$, see Fig.~\ref{fig:MFPT_memory}(a).
At high density, very strong memory becomes counterproductive: the superposition of many long-lived chemical trails leads to an over-suppression of trajectory crossing, which can hinder access to remaining unexplored regions and slow down the search.
Increasing $D_c$ or $\alpha_c$ weakens memory and partially restores access to these regions, at the cost of allowing some revisits.
The optimal search strategy therefore shifts toward intermediate memory strength, balancing revisit suppression against collective over-avoidance. The same collective over-avoidance mechanism also explains why the operational boundary defined by $\xi>1$ shifts toward larger $(D_c,\alpha_c)$ as the searcher density increases.

Within the strong-memory regime, the resulting collective speedup $\gamma_N$ is significantly larger than in the weak-memory regime, see Fig.~\ref{fig:MFPT_memory}(d).
Crucially, the origin of this speedup differs qualitatively from that discussed in Sec.~\ref{sec:low_memory}.
While efficient search at weak memory relies on spatial ordering and an effective partitioning of the search domain, spatial order remains negligible throughout the strong-memory regime [Fig.~\ref{fig:MFPT_memory}(e)].
The order parameter $\eta_{\mathcal A}$ stays small, and its global maximum appears only at larger $(D_c,\alpha_c)$, outside the strong-memory region, see Fig.~\ref{fig:MFPT_memory}(a). Instead, walkers in the strong-memory regime avoid not only their own long-lived chemical traces but also those produced by others, thereby reducing redundant exploration without inducing global alignment. We observe that path crossing is highly reduced and that the normalized crossing suppression $\chi$ is systematically enhanced as compared to the weak-memory regime, see Fig.~\ref{fig:MFPT_memory}(c). The resulting network of overlapping chemical traces forms a dynamically evolving constraint landscape, reminiscent of an entangled polymer melt, with walkers acting as mobile endpoints. It is this trail-mediated collective self-avoidance—rather than spatial order—that underlies the pronounced cooperative speedup observed at strong memory.

Taken together, these results establish the strong-memory regime as a distinct mode of collective search.
Long-lived but decaying chemical trails regulate exploration by suppressing revisits at both the single-particle and collective level.
Optimal search performance is achieved when memory is strong enough to encode past motion, yet sufficiently transient to avoid self-caging. This mechanism is fundamentally distinct from the order-driven enhancement operative in the weak-memory regime and identifies environmentally mediated memory, encoded in the self-generated chemical field, as an independent and powerful control parameter for collective search.

We note that other collective states known in chemotactic systems, such as traveling bands, can severely degrade search performance by concentrating exploration at the band front; a scaling analysis is provided in Appendix~\ref{app:bands}.

\section{Conclusions and outlook}
\label{sec:conclusion}

In this work we have shown that efficient collective search does not arise from spatial order alone, but from the mechanism by which redundant exploration is suppressed.
Using auto-chemorepulsive searchers as a minimal model that combines indirect interactions and memory, we identified two qualitatively distinct regimes of collective optimization, governed by the lifetime of self-generated chemical trails.

In the weak-memory regime, chemical cues are short-lived and primarily act to mediate repulsive interactions between agents.
Here, efficient search relies on a balance between directional persistence and spatial separation: moderate ordering distributes searchers across the domain while preserving sufficient mobility.
In this regime, excessive ordering or overly strong interactions reduce performance by constraining motion and promoting collective structures that do not contribute effectively to exploration.

In contrast, the strong-memory regime is controlled by long-lived and localized chemical trails that encode information about past trajectories.
Search efficiency is then enhanced not by global spatial order, which remains negligible, but by trail-mediated self-avoidance.
At both the single-agent and collective level, long-lived trails suppress revisits and strongly reduce long search times, leading to compressed-exponential survival statistics and pronounced cooperative speedup.
Importantly, optimal search is achieved at finite memory strength: permanently persistent trails induce self-caging and fragment the accessible search space, while moderate decay allows re-entry into previously enclosed regions and maintains global connectivity.

These results demonstrate that collective search efficiency can be regulated by qualitatively different physical mechanisms depending on memory strength.
Weak memory favors order-driven domain sharing, whereas strong memory gives rise to collective self-avoidance without alignment or large-scale ordering.
This distinction highlights memory as an independent and powerful control parameter for collective search, rather than a secondary correction to persistence or interaction strength.

Beyond auto-chemotactic systems, our findings point to a broader principle: interactions enhance collective search only insofar as they suppress redundant exploration without overconstraining motion.
The similarity between chemotactic repulsion, soft pairwise interactions, and even enforced domain partitioning underscores that the microscopic origin of avoidance is less important than its effective impact on exploration dynamics.

Several open directions remain.
Alternative forms of self-generated gradients, including attractant sinks \cite{SelfgenGrad1,SelfgenGrad2,fu2018nc} or chemokinetic responses \cite{zusman2007chemosensory,dorsogna2003pre}, may realize different trade-offs between memory and mobility.
More complex cost functions that account for signal production or energetic constraints could modify optimal strategies \cite{meyer2025optimal}, particularly in dense or competitive environments.
Finally, extending these ideas to adaptive or learning agents 
\cite{Volpe-MI,IntelligentMatter,Golestanian-MI,Liebchen-MI},
may help bridge stochastic search theory, active-matter physics, and collective decision-making in biological and artificial systems.

Taken together, our work establishes auto-chemotactic repulsion as a minimal and versatile framework for studying collective search, and reveals how tuning memory and interactions can switch the dominant mechanism by which groups explore space efficiently.

\section{Acknowledgments}
We acknowledge financial support from DFG through the Collaborative Research Center SFB 1027. This work was supported by the Medical Research Council (Grant No. MR/Y003845/1). Parts of the manuscript were revised and edited using ChatGPT (OpenAI) to improve clarity and expression. We thank Reza Shaebani for helpful discussions.

\appendix

\section{Off-lattice model details}
\label{app:models}

Particles are disks of radius $a$ interacting through a harmonic repulsion $V(r)=\frac{k}{2}(2a-r)^2$ for $r \le 2a$, with stiffness $k$. The force on particle $i$ due to particle $j$ is ${\bf f}_{ij}=-\nabla_{\mathbf{r}_i}V(r_{ij})$, where $r_{ij}=|{\bf r}_i(t)-{\bf r}_j(t)|$ is the distance between particles $i$ and $j$. 

The source term in Eq.~\ref{eq:chem_diffusion} is regularized using a Gaussian profile centered at $\mathbf{r}_i-a\mathbf{e}_i$ to reduce trivial self-interaction, and the local gradient in Eq.~\ref{eq:chemotaxis} is sampled at $\mathbf{r}_i+a\mathbf{e}_i$.

Rotational noise $\xi_i(t)$ in Eq.~\ref{eq:chemotaxis} is Gaussian white noise with $\langle\xi_i(t)\rangle=0$ and $\langle \xi_i(t)\xi_j(t')\rangle=\delta_{ij}\delta(t-t')$. We assume that the translational and rotational friction constants correspond to those of a sphere with radius $a$, i.e. $\gamma_r/\gamma_t=4/3\,a^2$. 

We use a simulation box of size $L/a=100$ with periodic boundary conditions. The system is first equilibrated before starting any measurements.

%NEW

\section{Stationary concentration profile}
\label{app:concentration_profile}

Consider a two-dimensional diffusion equation with a single source moving at constant speed $v_0$:
\begin{equation}
\frac{\partial c}{\partial t}(\mathbf r,t)
=
D_c \nabla^2 c(\mathbf r,t)
- \alpha_c c(\mathbf r,t)
+ h_c \delta(x-v_0 t, y)\,.
\end{equation}
Introducing the comoving field
$\tilde c(x,y,t)=c(x+v_0 t,y,t)$,
the steady-state profile
$c_\infty(x,y)=\lim_{t\to\infty}\tilde c(x,y,t)$
satisfies
\begin{equation}
\nabla^2 c_\infty
+ 2\lambda_{\mathrm{adv}}^{-1}
\frac{\partial c_\infty}{\partial x}
- \lambda_{\mathrm{decay}}^{-2} c_\infty 
+ \frac{h_c}{D_c}\delta(x,y)
=0\,,
\end{equation}
where $\lambda_{\mathrm{adv}}=2D_c/v_0$ and $\lambda_{\mathrm{decay}}=\sqrt{D_c/\alpha_c}$. Taking the Fourier transform gives
\begin{equation}
\hat c_\infty(k_x,k_y)
=
\frac{h_c/D_c}
{k_x^2+k_y^2
-2 i k_x \lambda_{\mathrm{adv}}^{-1}
+\lambda_{\mathrm{decay}}^{-2}}\,.
\end{equation}
Transforming back to real space yields
\begin{equation}
c_\infty(x,y)
=
\frac{h_c}{2\pi D_c}
e^{-x\lambda_{\mathrm{adv}}^{-1}}
K_0\!\left(
\kappa \sqrt{x^2+y^2}
\right),
\end{equation}
where $K_0$ is the modified Bessel function of the second kind and $\kappa^2=\lambda_{\mathrm{adv}}^{-2}+\lambda_{\mathrm{decay}}^{-2}$. For large distances $\kappa r \gg 1$, with $r=\sqrt{x^2+y^2}$, the asymptotic form
\[
K_0(u)\simeq
\sqrt{\frac{\pi}{2u}}\,e^{-u}
\qquad (u\to\infty)
\]
gives along the $x$-axis
\begin{equation}
c_\infty(x,0)
\approx
\frac{h_c}{D_c\sqrt{8\pi\kappa |x|}}
\begin{cases}
e^{-(\kappa-\lambda_{\mathrm{adv}}^{-1})|x|} &\text{if}\; x<0\\[4pt]
e^{-(\kappa+\lambda_{\mathrm{adv}}^{-1})|x|} &\text{if}\; x>0\,.
\end{cases}
\end{equation}
The decay length therefore differs ahead and behind the particle, leading to the fore–aft asymmetry discussed in the main text.

For $\kappa\gg\lambda_{\mathrm{adv}}^{-1}$ or, equivalently, for $\lambda_{\mathrm{decay}}\ll\lambda_{\mathrm{adv}}$, the profile becomes isotropic,
\begin{equation}
c_\infty(r)
\sim
\frac{h_c}{D_c\sqrt{4\pi\kappa r}}
e^{-\kappa r},
\end{equation}
recovering the two-dimensional Yukawa form.

\section{Order--persistence relation in the weak-memory regime}
\label{app:order_persistence}

\begin{figure}
    \centering
    \includegraphics[width=.85\linewidth]{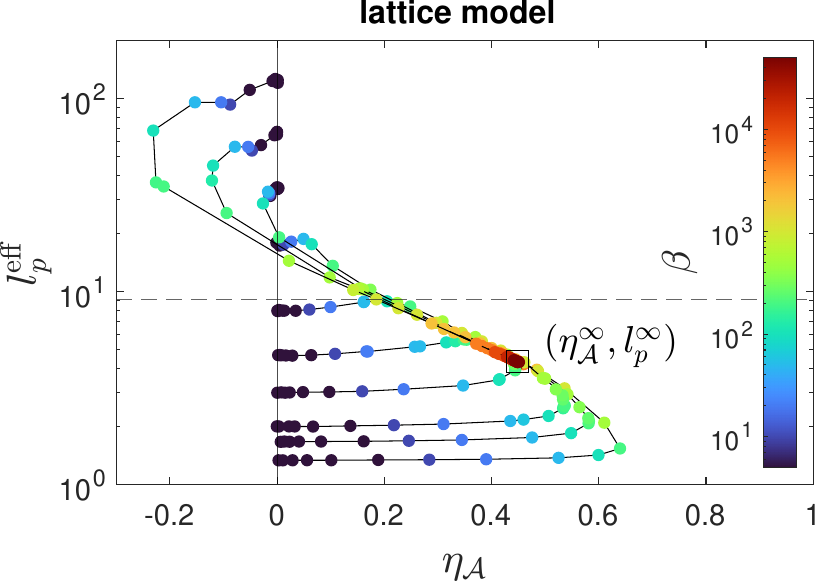}
    \caption{Effective persistence length $l_p^\mathrm{eff}$ versus spatial order $\eta_\mathcal{A}$ for fixed density $\varphi=0.012$ of the auto-chemotactic lattice model. For each bare persistence length $l_p$, increasing $\beta$ traces out a distinct trajectory in the $(\eta_\mathcal{A}, l_p^\mathrm{eff})$ plane that converges to a common strong-coupling state $(\eta_\mathcal{A}^\infty, l_p^\infty)$.}
    \label{fig:lpeff_vs_order}
\end{figure}

In the weak-memory regime, spatial order and effective persistence cannot be tuned independently. Varying the bare persistence $l_p$ and the chemotactic coupling $\beta$ simultaneously modifies both the effective persistence length $l_p^{\mathrm{eff}}$ and the spatial order parameter $\eta_\mathcal{A}$. Figure~\ref{fig:lpeff_vs_order} shows the resulting trajectories in the $(\eta_\mathcal{A}, l_p^{\mathrm{eff}})$ plane at fixed density. Two regimes are observed. For moderate $\beta$, spatial order increases while $l_p^{\mathrm{eff}}\approx l_p$ remains nearly unchanged, indicating that repulsive separation dominates. For larger $\beta$, the data collapse onto a boundary curve, revealing a trade-off: for a given $l_p^{\mathrm{eff}}$, spatial order is bounded, and conversely, high spatial order cannot be achieved at arbitrarily large persistence. This constraint originates from the finite interaction range $\lambda_{\mathrm{decay}}$. Long-range velocity correlations would be required to achieve both high persistence and strong ordering simultaneously, but chemotactic alignment is limited when $\lambda_{\mathrm{decay}}$ is shorter than the persistence length or the typical interparticle spacing $1/\sqrt{\varphi}$. As a result, the maximal attainable order decreases with increasing $l_p$. In the strong-coupling limit $\beta\to\infty$, for all values of $l_p$, the system converges to a common point $(\eta_\mathcal{A}^\infty, l_p^\infty)$. In this regime, particles follow the steepest descent of the concentration field and move ballistically until redirected by collisions. The typical trajectory length is therefore set by a mean free path $l_{\mathrm{MFP}}=1/(\sqrt{8}\,\varphi d)$, where $d\simeq\lambda_{\mathrm{decay}}/3$ is the effective interaction diameter. This confirms that in the strong-coupling regime, collective persistence is collision-limited.

\section{Identification of memory regimes from first-passage statistics}
\label{app:regimes}

\begin{figure}
    \centering
    \includegraphics[width=.8\linewidth]{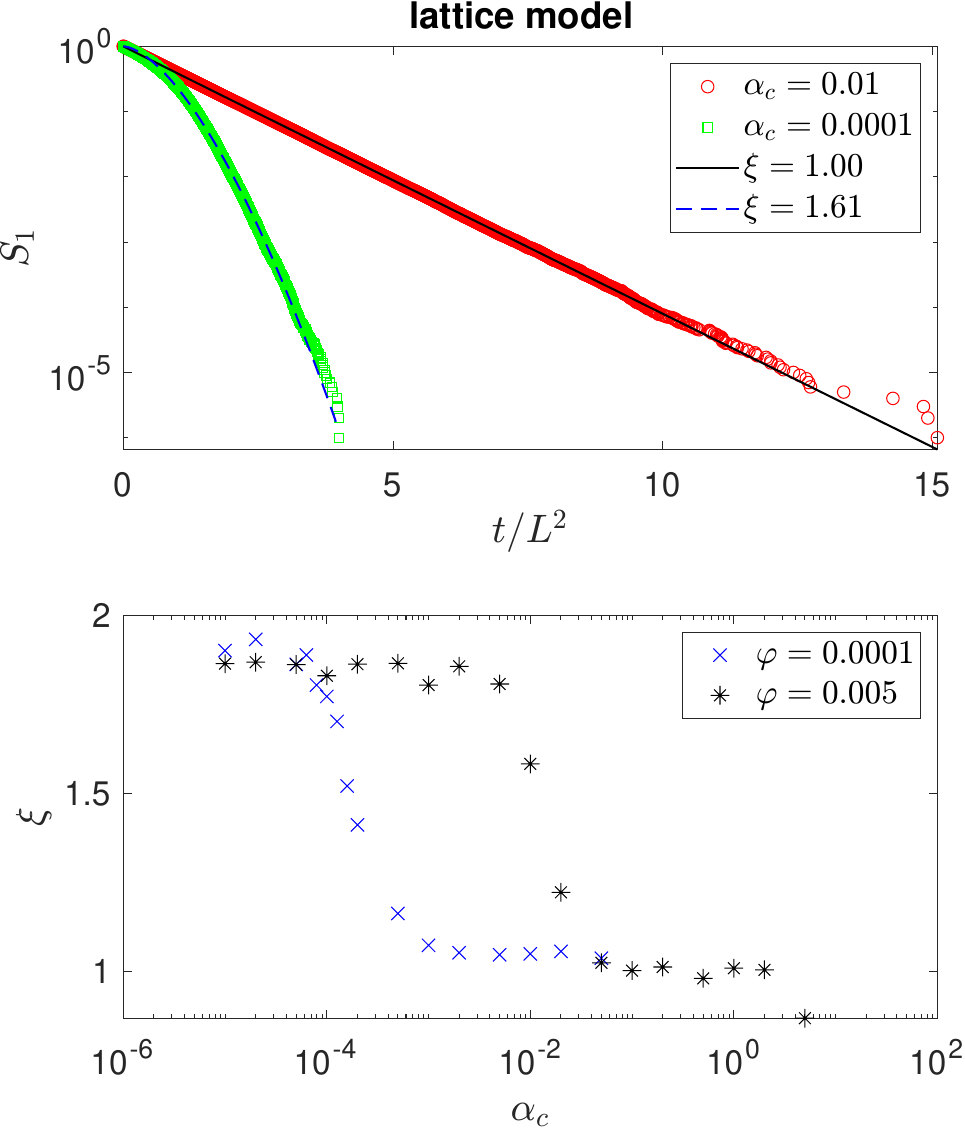}
    \caption{(a) Single-particle survival probability $S_1(t)$ in the lattice model for $\varphi=10^{-4}$ and $D_c=10^{-4}$ at two values of the decay rate $\alpha_c$. Lines show stretched-exponential fits $S_1(t)=\exp[-(t/T)^\xi]$. (b) Stretching exponent $\xi$ as a function of $\alpha_c$ for two densities $\varphi$ at fixed $D_c=10^{-4}$. The crossover from $\xi\approx1$ to $\xi>1$ marks the transition from weak- to strong-memory behavior.}
    \label{fig:xi}
\end{figure}

To identify the crossover between weak- and strong-memory behavior, we fitted the survival probability $S_N(t)$ to the stretched-exponential form $S_N(t)=\exp[-(t/T)^\xi]$. Figure~\ref{fig:xi}(a) shows representative survival probabilities for a single auto-chemotactic random walker ($N=1$) together with the corresponding fits. For large $\alpha_c$, the stretching exponent remains close to $\xi\simeq1$, corresponding to approximately exponential first-passage statistics. As $\alpha_c$ decreases, $\xi$ increases above unity, indicating compressed-exponential survival. The resulting dependence of $\xi$ on $\alpha_c$ for two different densities $\varphi$ is shown in Fig.~\ref{fig:xi}(b). We use the crossover from $\xi\approx1$ to $\xi>1$ to identify the onset of the strong-memory regime.

\section{Finite-memory self-avoidance: snake model}
\label{app:snake}

To disentangle the effect of memory from the specific chemotactic interaction mechanism, we introduce a minimal lattice model in which self-avoidance is imposed explicitly over a finite temporal window.
This ``snake'' model serves as a conceptual reference that isolates the role of memory while preserving the essential trade-off between revisit suppression and self-caging discussed in Sec.~\ref{sec:strong_memory}.

We consider a single walker on a two-dimensional square lattice of size $L\times L$ with periodic boundaries boundaries.
Each lattice site $i$ carries a clock variable $\tau_i$ that records the time elapsed since the site was last visited.
At each time step, the walker attempts to jump to a neighboring site $j$ with probability
\begin{equation}
p_j \propto \exp\!\left[-\beta\,\Theta(\tau_j - l_{\mathrm{snake}})\right],
\end{equation}
where $l_{\mathrm{snake}}$ defines the memory length and $\Theta$ is the Heaviside function.
Sites visited within the last $l_{\mathrm{snake}}$ steps are therefore disfavored, while older sites are revisited without penalty.
After each move, the clock field is updated as $\tau_i \to \tau_i + 1$ on all sites except the current position, where $\tau_i$ is reset to zero.

For $l_{\mathrm{snake}}=0$, the dynamics reduces to an unbiased random walk.
In the opposite limit $l_{\mathrm{snake}}\to\infty$, revisits are permanently forbidden and the trajectory approaches that of a true self-avoiding walk.
The parameter $l_{\mathrm{snake}}$ thus provides a direct control over the memory depth of the searcher.

Figure~\ref{fig:snake_mfpt} shows the mean first-passage time (MFPT) to a randomly placed target as a function of the normalized memory length $l_{\mathrm{snake}}/L^2$ for several system sizes.
In all cases, the MFPT exhibits a pronounced minimum at finite memory.
When rescaled by the system area, the optimal memory length collapses as
\begin{equation}
l_{\mathrm{snake}}^\ast \sim L^2,
\end{equation}
corresponding to the typical time required to explore the domain once.

For memory lengths shorter than $l_{\mathrm{snake}}^\ast$, frequent revisits dominate and search efficiency is reduced.
For longer memory lengths, self-caging effects emerge as permanently excluded regions fragment the accessible space, again increasing the MFPT.
This minimal model therefore reproduces the same qualitative trade-off observed in auto-chemotactic search: efficient exploration requires memory that is strong enough to suppress redundant revisits, yet sufficiently transient to avoid topological trapping.

Importantly, the snake model contains neither chemotactic interactions nor collective effects.
Its agreement with the auto-chemotactic results demonstrates that the existence of an optimal, finite memory length is a generic consequence of constrained exploration dynamics rather than a model-specific feature.

\section{Inefficient search by bands}
\label{app:bands}

\begin{figure}
    \centering
    \includegraphics[width=.95\linewidth]{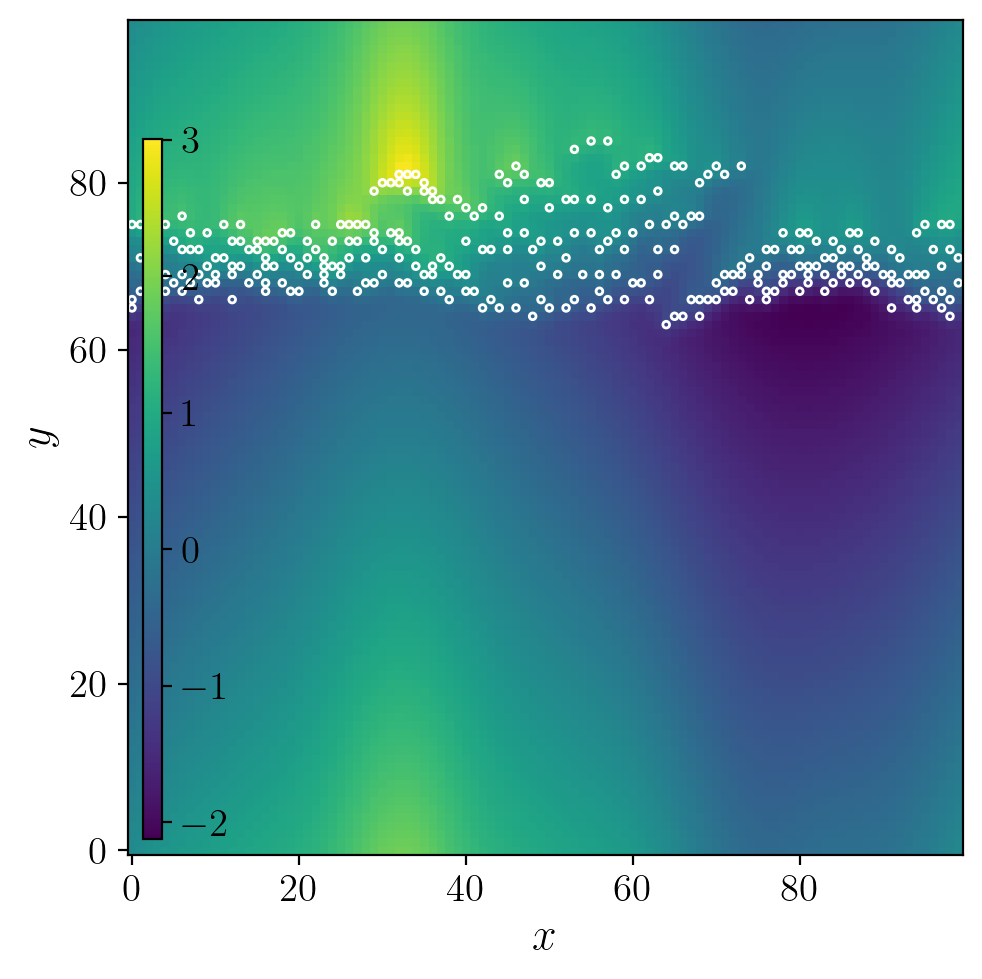}
    \caption{Example of a band formed by auto-chemotactic \adam{lattice} random walkers. The background color shows the chemical concentration field, while white circles mark the walkers. The band spans the system horizontally and propagates downward at approximately constant speed.}
    \label{fig:band_snapshot}
\end{figure}

Auto-chemotactic particles are known to display a rich collective phenomenology beyond homogeneous states
\cite{liebchen1,liebchen2,liebchen3,kranz2019trail, hokmabad2022chemotactic, mokhtari2022spontaneous, barbier2022self, moreno2022single, meyer2023alignment}.
In particular, auto-chemorepulsive interactions can induce effective alignment, which at sufficiently high particle density and for slowly diffusing chemical cues leads to the spontaneous formation of traveling bands
\cite{liebchen1,meyer2023alignment}. 
Such banded states are also observed in our lattice simulations, see Fig.~\ref{fig:band_snapshot}.
A typical band spans the system along one direction and propagates ballistically in the transverse direction, often containing a large fraction of all searchers.

While band formation represents a robust emergent pattern of chemotactic interactions, it is strongly disadvantageous for collective search.
Because only walkers located at the leading edge of the band encounter unexplored territory, particles in the bulk repeatedly rescan regions that have already been visited.
As a result, the search efficiency is effectively controlled by the motion of the band front alone.

This observation allows for a simple estimate of the mean first-passage time (MFPT) in the banded state.
For a system of linear size $L$ in which a band propagates ballistically at speed $v_0$, the MFPT scales as
\begin{equation}
\bar{T}_{\mathrm{band}} \simeq \frac{L}{2v_0},
\end{equation}
independent of the total number of searchers $N$.
Here, dimensional units are used for clarity.

For comparison, $N$ independent persistent searchers homogeneously distributed across the domain achieve an MFPT
\begin{equation}
\bar{T}_{\mathrm{hom}} \simeq \frac{L^2}{4Nv_0 a},
\end{equation}
where $a$ is the particle radius.
This expression applies in the ballistic regime $l_p \gg a$ and represents the optimal scaling for independent active Brownian particles.
The factor $a$ reflects the effective transverse width $2a$ swept out by each particle.

Comparing the two estimates yields
\begin{equation}
\frac{\bar{T}_{\mathrm{hom}}}{\bar{T}_{\mathrm{band}}}
= \frac{L}{2Na},
\label{eq:band_vs_hom}
\end{equation}
which is smaller than unity as soon as the number of searchers exceeds the minimal number required to span the system.
Physically, this reflects the fact that in a homogeneous population all particles contribute to parallel exploration, whereas in a banded state only the leading edge explores new territory.

For less persistent walkers, the corresponding MFPT estimate for homogeneous search acquires logarithmic corrections of order $\ln L$, but the qualitative conclusion remains unchanged. In that case one finds $\bar{T}_{\mathrm{hom}}/\bar{T}_{\mathrm{band}} \sim L/(N l_p)$ for $l_p \ll L$. An analogous scaling argument can be formulated for persistent random walkers.

In chemotactic systems, bands typically contain far more particles than the minimal number required to span the system.
Equation~(\ref{eq:band_vs_hom}) therefore implies that the MFPT in the banded state is substantially larger than in the homogeneous state.
We thus conclude that, despite being a common outcome of chemotactic interactions, band formation suppresses the advantages of collective search by concentrating exploration at a single moving front.
Banding should therefore be regarded as a failure mode of collective search rather than an optimized strategy.

%\bibliography{autochemo_search_2}


\end{document}